# Tracking Filament Evolution in the Low Solar Corona using Remote-Sensing and In-situ Observations

**Manan Kocher***
Department of Climate and Space Sciences and Engineering, University of Michigan,
2455 Hayward St, Ann Arbor, MI 48109-2143, USA
mkocher@umich.edu | [734]-355-4944

**Enrico Landi**
Department of Climate and Space Sciences and Engineering, University of Michigan,
2455 Hayward St, Ann Arbor, MI 48109-2143, USA | elandi@umich.edu

**Susan. T. Lepri**
Department of Climate and Space Sciences and Engineering, University of Michigan,
2455 Hayward St, Ann Arbor, MI 48109-2143, USA | slepri@umich.edu

# Abstract

In the present work, we analyze a filament eruption associated with an ICME that arrived at L1 on August $5^{th}$, 2011. In multi-wavelength *SDO/AIA* images, three plasma parcels within the filament were tracked at high-cadence along the solar corona. A novel absorption diagnostic technique was applied to the filament material travelling along the three chosen trajectories to compute the column density and temperature evolution in time. Kinematics of the filamentary material were estimated using *STEREO/EUVI* and *STEREO/COR1* observations. The Michigan Ionization Code used inputs of these density, temperature, and speed profiles for the computation of ionization profiles of the filament plasma. Based on these measurements we conclude the core plasma was in near ionization equilibrium, and the ionization states were not frozen-in at the altitudes where they were visible in absorption in *AIA* images. Additionally, we report that the filament plasma was heterogeneous, and the filamentary material was continuously heated as it expanded in the low solar corona.

# Keywords

Sun: corona, Sun: coronal mass ejections (CMEs), Sun: filaments, prominences
(Sun:) solar wind, Sun: UV radiation, (Sun:) solar-terrestrial relations

# 1. INTRODUCTION

Coronal Mass Ejections (CMEs) are large-scale solar transients that deposit vast amounts of plasma and magnetic energy into the heliosphere. Our scientific interest in CME physics spans: determining their geo-effectiveness at Earth to protect our technological infrastructure, understanding their spatial and kinematic evolution in the heliosphere based on how they interact with the ambient solar wind, and dissecting the dynamic processes that lead to their formation at their solar source. In-situ measurements from spacecraft such as the *Advanced Composition Explorer (ACE), WIND,* and *Ulysses*, and remote-sensing observations from spacecraft such as *SOHO, STEREO,* and *Solar Dynamic Observatory (SDO)* have vastly contributed to our current understanding of CMEs. Recent studies such as Gruesbeck et al. (2011), Landi et al. (2012), and Rodkin et al. (2017) have discussed and implemented techniques that use combinations of in-situ and remote-sensing measurements to provide additional insight into CME dynamics at their source and through their heliospheric evolution. In-situ observations of Interplanetary Coronal Mass Ejections (ICMEs, heliospheric counterparts of CMEs) tell us about the state of the plasma after the freeze-in process (Geiss et al. 1995), while remote-sensing observations of the CMEs contribute to our understanding of CMEs immediately after eruption and near their coronal source.

White-light coronagraph observations of CMEs have commonly revealed distinct 3-part structures: a bright and dense front, trailed by a low-density cavity which contains a relatively bright, high density core (Howard et al. 1985; Illing and Hundhausen, 1985; Hundhausen, 1987, 1999). The energetics of a CME are best studied if we know the varying plasma density, temperature, and dynamics of these various components. Extreme Ultraviolet (EUV) and X-ray observations of CMEs are frequently used to view and diagnose the different CME features low in the solar corona. The Extreme Ultraviolet Imager (EUVI, Howard et al. 2008) on board the twin STEREO spacecraft (Kaiser et al. 2008) and the Advanced Imaging Assembly (*AIA*) (Lemen et al. 2012), onboard *SDO* (Pesnell et al. 2012) provide multi-wavelength observations in EUV wavebands, allowing for detailed and high-cadence diagnostics. Often observed in EUV images, filaments are large regions of dense and cold gas held in place by magnetic fields. Upon eruption, they frequently form the dense core of the CME (House et al. 1981; Webb and Hundhausen, 1987; Gilbert et al. 2000; Zhang et al. 2001; Gopalswamy et al. 2003). Since energetics of the CME core can provide credible estimates of the energetics of the entire CME itself (Landi et al. 2010), filament/prominence observations are highly valuable to our understanding of processes involved in the onset of CMEs. The current investigation focuses on the heating and acceleration processes in an erupting filament.

Researchers have previously produced reconstructions of eruptive phenomena in simulated solar coronas using mathematical models (see review in Chen 2011). The success of these models relies on a combination of rigorous MHD theory and reliable observations. An understanding of spectroscopic observations of CMEs and their associated phenomena, such as filaments and flares, is critical to replicate the observed CME evolution in the models using MHD theory. Another usage of spectroscopic observations is to estimate CME thermodynamic and kinematic characteristics that can serve as effective boundary conditions to different models of CME evolution using diagnostic techniques. Studies such as Gilbert et al. (2005), Gilbert et al. (2011), Landi & Reale (2013), Landi & Miralles (2014), Hannah & Kontar (2012, 2013), and Williams et al. (2013) developed techniques that measure properties of transient plasmas utilizing dark absorption features or bright emission features from EUV and X-ray images. The absorption diagnostic technique outlined by Landi & Reale (2013) was utilized to produce the results described in the current investigation.

A recent study by Lee et al. (2017) demonstrated some of the abovementioned diagnostic techniques in their investigation of an erupting prominence and loops associated with a January 25$^{th}$, 2012 CME observed by *SDO/AIA* first as absorption features, which then change into emission features as the plasma heated up. They used a polychromatic diagnostic technique to compute column densities of the absorbing plasma, performed the Differential Emission Measure (DEM) analysis to compute column densities and temperatures of the emitting plasma, estimated the mass of the erupting loops and prominence, and computed the energetics and heating of these erupting features. Another study by Rodkin et al. (2017) performed DEM diagnostics on multi-wavelength *SDO/AIA* observations of a series of solar wind transients to derive averaged electron temperatures and densities of the CME plasma structures under consideration at five different times. However, in order to utilize derived CME plasma parameters to constraint models and validate the heating and acceleration processes they account for, these parameters need to be derived over longer periods of time at higher cadence while ensuring the same transient plasma is tracked. In the current investigation, we introduce a novel procedure to follow plasma parcels within an erupting filament in high-cadence *AIA* images.

As CMEs evolve through the outer corona they can be tracked in coronagraph images and heliospheric images from the STEREO and SOHO (Domingo et al. 1995) missions. In addition to the evolution of their thermal and kinetic properties, the evolution of various CME component ionic compositions can be sensitive indicators of the thermal environment in which the plasma existed in the low solar corona. The composition of plasma parcels is controlled by ionization and recombination processes, which are in turn modulated by the local electron density, temperature, and speed of the parcel. Since the electron density of the plasma drops due to expansion as it travels to farther distances above the solar surface, the atomic processes that shape the ionization distribution eventually shut down, resulting in the freeze-in of the plasma ionic composition within 1-5 solar radii (Hundhausen et al. 1968, Hundhasusen, 1972, Bame et al. 1974, Buergi & Geiss, 1986, Geiss et al. 1995). It is critical to note that the freeze-in region would vary according to the species and the varying plasma properties of the different CME components. This frozen-in composition is then observed by in-situ instruments such as the Solar Wind Ion Composition Spectrometer (*SWICS,* Gloeckler et al. 1998*)* on board *ACE* at L1. SWICS observations of ICMEs have revealed significant deviation from the composition of the ambient solar wind (e.g. Lepri et al. 2001, Zurbuchen & Richardson, 2006, Richardson & Cane, 2004, 2010, Kocher et al. 2017) allowing us to isolate them and draw inferences on the thermal environment in which the CME existed in the low solar corona before the freeze-in region. However, the composition of the CME before freeze-in cannot be determined directly from in-situ measurements because the plasma temperature inferred from in-situ measurements of the ion abundance ratio has little correlation with the actual temperature in the region before the freeze-in point (Landi et al. 2012). Gruesbeck et al. (2011) used in-situ ionic charge state measurements of a January 27$^{th}$, 2003 ICME from *ACE/SWICS* to constraint the early evolution of the CME plasma, focusing on the characteristic bi-modal nature of iron distribution with typical peaks at Fe$^{10+}$ and Fe$^{16+}$ (Lepri et al. 2001). By modeling the freeze-in condition of the CME using different temperature and density profiles, they concluded that the bi-modal iron charge distributions along with the observed carbon and oxygen distributions are best matched if the CME plasma underwent rapid heating followed by rapid expansion from a high initial density before freeze-in. Another effort by Lynch et al. (2011) computed two-dimensional spatial distributions of C, O, Si, and Fe for CME simulations initiated by flux cancellation (Reeves et al. 2010) and magnetic breakout (Lynch et al. 2008) processes, providing evidence of enhanced heavy ion charge states in the CME plasma formed as a consequence of flare heating during eruption initiation. CME science has yet to establish conclusively the nature of the composition of CMEs before freeze-in, the processes that shaped them, and the thermal environment in which they evolved in the low solar corona. In this investigation, we will model the composition within filament plasmas using thermal and kinematic observations of a filament eruption in the low solar corona.

We analyze a filament eruption associated with a CME from August $4^{th}$, 2011 that was geo-effective, tracked during the initial hours of its evolution in the low solar corona between 1.5-2.6 solar radii. We: (1) determine the evolution of the main physical parameters within the filament by following different plasma parcels in high-cadence EUV and white-light images, (2) discuss the time evolution of the partition of the thermal and kinetic energy of CME plasmas using the plasma parameters determined in the previous step, and (3) examine the evolution of charge-state composition of CME plasmas in the lower solar corona. This study attempts to build on existing groundwork for connecting in-situ and remote-sensing observations by addressing the question of how the dynamics and energetics of a geo-effective CME core evolved in the low solar corona. The rest of this report is organized into several sections. Section 2 describes the multi-instrument in-situ and remote-sensing observations of the filament eruption used to execute this study and how they were used to connect the same event in near-Earth space and at the Sun. Section 3 describes our technique to track the plasma evolution within the filament eruption in *AIA* images. Density and temperature determination of the filament plasma using absorption diagnostics are described in Section 4. Kinematics of the filament plasma in the low solar corona are presented in Section 5. The neutral hydrogen, proton, and electron density determination and results are presented in Section 6. In Section 7 we estimate the ionization history of the filament plasma using results from the previous sections as inputs to the Michigan Ionization Code (MIC, Landi et al. 2012a). Section 8 offers a discussion of our results followed by conclusion and future work associated with this investigation.

## 2. Multi-Instrument Observations of a Coronal Mass Ejection: *SWICS/ACE, SOHO/LASCO, STEREO/SECCHI, SDO/AIA*

This research is enabled in part by observations of geo-effective ICMEs from the composition sensors on board the Advanced Composition Explorer (ACE). The Richardson and Cane (2004, 2010) list (RC list) identifies ICMEs that arrived at Earth using plasma and magnetic field ICME signatures in the *ACE/SWICS, ACE/MAG* (MAGnetic Field Experiment), and *ACE/SWEPAM* (Solar Wind Electron, Proton and Alpha Monitor) measurements. Additionally, the RC list frequently identifies the corresponding *SOHO/LASCO* CME, allowing us to begin tracing the ICME back to the Sun. The instruments on the Solar and Heliospheric Observatory (*SOHO*) changed the landscape of CME science with its groundbreaking routine observations of Earth-bound transients. The Large Angle Spectrometric COronagraphs (*LASCO*, Brueckner et al., 1995) onboard *SOHO* observe the inner and outer solar corona between 1.1-32 solar radii – with C2 from 1.5-6 solar radii and C3 from 3.5-30 solar radii. The twin *STEREO* spacecraft further enhanced our ability to study CMEs by placing identical Sun-Earth Connection Coronal & Heliospheric Investigation (*SECCHI*) fleet of instruments onboard the two spacecraft, allowing for multi-perspective viewing from 1-15 solar radii and 3D reconstruction of ICMEs/CMEs. The overlapping fields of view of the *SECCHI* instruments enabled us to connect the same structure as it expanded away from the Sun and crossed the different instruments fields of view. The Extreme Ultraviolet Imagers (*EUVI*) instruments on both STEREO spacecraft observe the solar corona below 1.7 solar radii in 171 Å, 195 Å, 284 Å, and 304 Å channels. Additionally, four coronagraphs part of the *STEREO/SECCHI* instrument suite allow us to analyze the faint emissions of the solar corona by blocking light from the brighter solar disk. COR 1 is a Lyot coronagraph that observes the solar corona between 1.4-4 solar radii, and COR 2 is a Lyot coronagraph that observes the solar corona between 2.5-15 solar radii. Furthermore, the launch of the Solar Dynamic Observatory (*SDO*) gave us pioneering tools to study solar sources with the Atmospheric Imaging Assembly (*AIA*). *SDO/AIA* allow us to analyze the solar corona with its high-cadence, multi-wavelength full Sun EUV observations. The CME under scrutiny in this study was chosen based on its visibility in observations from a majority of these instruments.

Our analysis utilized the case of a filament eruption at the Sun associated with an ICME that arrived at Lagrange 1 (L1) point on August 5$^{th}$, 2011. The Richardson and Cane (2004, 2010) ICME list (RC list) identified the arrival of this intense storm on August 5$^{th}$, 2011 at 17:51 UT. We traced this geo-effective ICME back to the solar corona using a combination of multi-instrument remote-sensing and imaging observations to perform time-evolution diagnostics of the properties of the eruption during the initial hours of its journey while it was still in the low solar corona. At the time of the CME eruption, *SDO* was operational, allowing us to use the high-cadence measurements from *AIA*. Furthermore, the twin *STEREO* spacecraft were in near-quadrature with *SDO*; that is, the same CME observed by *STEREO* at the solar limb was observed by *SDO* close to the disk center, minimizing the uncertainty in the 3D reconstructions of the CME plasma.

The RC list pointed us in the direction of the *LASCO* CME associated with our ICME under scrutiny. In the C2 coronagraph the CME eruption was seen starting at ~04:12 UT on August 4$^{th}$, 2011 shortly after which a partial halo enveloped the solar disk off the west limb (Figure 1(h)). The C3 coronagraph observes the CME eruption starting at ~04:42 UT after which the CME partially enveloped the solar disk (Figure 1(e)).

The ICME of interest was observed in *EUVI, COR1*, and *COR2* instruments on *STEREO A* and *STEREO B*. *STEREO A/EUVI* observed a flare starting ~03:46 UT followed by the filament eruption at ~04:06 UT, shortly after which the front of the eruption was outside the field of view of the instrument. The flare was not observed in *STEREO B/EUVI* images but the filament eruption observations were similar to that of *STEREO A/EUVI*. Figure 1(c) shows the *STEREO A/EUVI* image of the Sun in 304 Å at 04:36:15 UT, and Figure 1(a) shows the *STEREO B/EUVI* image of the Sun in 304 Å at 04:36:55 UT. A distinct front was seen in *COR1* and *COR2* coronagraphs on both *STEREO* spacecraft starting in images at ~04:10 UT followed by a bright jet-like filament eruption that formed the core of the CME further out in the corona.

In order to uniquely determine the source region of this eruption measured in-situ and by instruments onboard *SOHO* and *STEREO*, we used the following: (1) X-ray observations from the Geostationary Operational Environmental Satellite (*GOES*), (2) the Global High-Resolution H-alpha Network (*GHN*), and (3) multi-wavelength observations from *SDO/AIA*.
(1) X-ray observations from *GOES* identified an M-class flare from active region 11261 shortly before a CME erupted at 4:10 UT. This active region was also close to the Sun center indicating that a CME eruption here, associated with the highly energetic flare (also observed in *STEREO A/EUVI*) could possibly have been geo-effective.

(2) H-alpha images from 18:30:36 UT on August 3rd, 2011 and 08:04:50 UT on August 4th, 2011 indicated the disappearance of filaments near active region 11261 between the times the images were taken.

(3) Shortly after the flare eruption, *SDO/AIA* time-series images confirmed the eruption of a filament in the vicinity of the AR 11261 at ~04:10 UT. We anticipate the flare and filament eruption resulted from significant reorganizing and shearing of the magnetic topology associated with magnetic reconnection in the low solar corona. The filament eruption was clearly visible as a dark absorption feature in *AIA*'s high cadence data during in the initial fifty minutes of its eruption from AR 11261. Figure 1(b) shows a snapshot of this eruption at ~04:36 UT in the 304 Å channel.

Using the instruments described above, we were able to track the August 5th, 2011 ICME disturbance observed by *ACE* at L1 at 17:51 UT back to a filament eruption near active region 11261 on August 4th, 2011 at ~04:10 UT and observe it during its evolution. As an additional test, we used the WSA-ENLIL (Jian et al. 2015), a 3D time-dependent MHD model that predicts the arrival of Earth-directed CMEs capable of inducing geomagnetic storms by mapping the flow evolution out to Earth. It uses white-light images from the *LASCO* and *STEREO* coronagraphs. This model traced the origins of the August 5th, 2011 ICME at L1 to a solar event on August 4th, 2011 at ~04:10 UT, confirming our multi-instrument analysis described above.

In the following sections, we describe our use of multi-wavelength images from *SDO/AIA* to perform diagnostics as we followed the progress of the filament eruption associated with this CME eruption in the low solar corona.

## 3. Time-Dependent Tracking of the Plasma within the Filament Eruption in the Low Solar Corona:

The first step in determining the time evolution of our erupting filament's properties is charting possible paths travelled by the plasma within the eruption, while attempting to ensure the 'same plasma' is tracked over time. *AIA* images of the eruption from 03:45 UT – 05:15 UT, August 4th, 2011 were used to carry out this investigation. They were secured from the Solar Software (SSW) cutout service available online[1] in 171 Å, 193 Å, 211 Å, 304 Å, and 335 Å wavelength channels. The justification of the choice of these five wavelength channels is presented in Section 4.1. The filament was observed as an expanding dark absorption feature in the north-western hemisphere of the Sun near active region 11261. The multi-wavelength Flexible Image Transport System (*fits*) files were prepared for analysis using two steps. (1) The Automatic Exposure Control (AEC) effect due to the flare that preceded the filament eruption resulted in a reduced dataset with 24 second cadence during certain periods instead of the expected 12 seconds in some of the channels. Due to these data gaps, the data sets in all wavelength channels were reduced so that each image in one channel corresponded to the image at the closest time-step in all the remaining channels. This gave us *fits* files for the eruption in five wavelength regions with time cadences between 12 and 24 seconds. (2) The aia_prep.pro[2] IDL routine was then used to convert all *fits* files from level 1.0 to level 1.5, performing image rotation, translation, and scaling and updating the header information.

The method used to determine paths of plasma as the filament erupted using the high-cadence processed *AIA* images is described here. First, a 30x30 pixel parcel was picked within the absorbing plasma at 04:10 UT (Fig 2, left). This is the first time-stamp when the eruption was visible in absorption. After 04:10 UT, the absorbing plasma expanded outwards as it was accelerated through the inner corona. Hence, the plasma within the box in the left panel of Fig 2 could be assumed to have traveled along multiple paths. We strategically stepped this small pixel box 'outward' at each of the 102 time-steps through 05:00 UT (Fig 2, right) along three paths, loosely following specific features within the filament eruption that were easier to discern than others. Care was taken to ensure the chosen boxes contained absorbing plasma (i.e., remained dark) at each time-step under analysis. These three trajectories are shown in Figure 2 across three different times during the evolution of the eruption. Paths 1, 2, and 3 represent three of the possible paths a plasma parcel travelled along as the filament expanded outwards from the site of eruption and spread through the inner corona. While this tracing technique is not precise due to the complications of 3D structures viewed in 2D, it allowed for the elimination of averaging effects that would have resulted from performing diagnostics on large portions of the erupting plasma, and considering three possible paths allowed for a thorough comparative analysis on different parts of the filament.

## 4. Density & Temperature Diagnostics of the Filament Plasma in the Low Solar Corona:

### 4.1. Absorption Diagnostic Technique:

We used the absorption diagnostic technique of Landi & Reale (2013) to measure the column density and electron temperature of the absorbing plasma within the erupting filament. This technique capitalized on the EUV absorption properties of the filament material through bound-free transitions because in four wavelength ranges from 100-1000 Å [100-228 Å, 228-504 Å, 504-912Å, 912-1000Å] the absorption coefficient has a unique temperature dependency depending on which ions ($H^{1+}$, $He^{2+}$, $He^{1+}$) absorb in each of those wavelength regions. The temperature diagnostics from absorption comes from the fact that the relative abundance of those ions depend on the electron temperature; thus, it can be applied only to plasmas hotter than 15,000 K (below which H and He are almost entirely neutral) and cooler than ~100,000 K, where the abundance of $H^{1+}$, $He^{2+}$, $He^{1+}$ are too small to provide significant absorption. Our ability to uniquely observe the filament eruption as distinct absorption features in 171Å, 193 Å, 211 Å, 304 Å, and 335 Å channels of *SDO/AIA* allowed us to implement this absorption diagnostic technique on these high-cadence images since these five passbands were within two of the four wavelength ranges. The amount of radiation absorbed by a plasma parcel can be computed using Beer's law which states that:

$F_{abs} = F_{inc}\, e^{-\tau}$ Equation 1

Where, $F_{inc}$ is the intensity of incident radiation on an absorbing plasma parcel, $F_{abs}$ is the intensity measured after absorption, and $\tau$ is the optical depth. Adapted from Landi and Reale (2013), Figure 3 shows a simplified geometry of an absorbing filament plasma parcel in the low solar corona. Here, $F_a$ and $F_b$ are measured EUV intensity values observed along two lines of sight: '*a*' intercepts the absorbing filament plasma, and '*b*' can be envisioned as the intensity at the location in the absence of the filament plasma. The line of sight can be divided into three regions: finite depth of the filament, region behind the filament (background), and region between the observer and the filament (foreground). In order to simplify the number of unknowns, Landi and Reale (2013) outlined three different cases to execute the diagnostic technique based on reasonable approximations on the absorbing plasma. In testing the three cases, we estimated the scale height of coronal emission by taking a 30-pixel high horizontal cut in all 102 *AIA* images used between 04:10 UT – 5:00 UT in the northern hemisphere near the western limb. The distance at which the intensity dropped by a factor of 1/*e* from the edge of the solar disk was approximately 0.06 solar radii with insignificant standard deviation. Since the filament eruption was already at 1.7 solar radii by 04:10 UT (determined using *STEREO A* observations, Section 5*)* we followed the assumptions in the case where the absorbing material (filament) is located at altitudes significantly greater than the scale height of the coronal emission in the solar atmosphere. This allowed us to neglect the foreground emission along the line of sight compared to the background emission i.e., $F_3 \ll F_1$. Estimates of the line of sight depth of the absorbing plasma (Section 6) show that $F_2 \ll F_1$ could also be assumed. Furthermore, Equation 1 implicitly assumes that the filament itself is not emitting at the wavelengths used for diagnostics. This is true for all *AIA* channels we considered except, in principle, the 304 Å channel, which is dominated by the strong HeII 303.7 Å transition which could contribute to observed counts. In this work, we assume that for the 304 Å channel HeII emission is negligible, but discuss this assumption and its limitations in Section 8. Under these assumptions, Equation 1 can be used to estimate the optical depth from the measured values of $F_a$ and $F_b$ as follows:

$\tau = -\ln(F_a/F_b)$                                 Equation 2

where $F_2 \ll F_1$ and $F_3 \ll F_1$, based on our corroborated geometric assumptions.

Landi and Reale (2013) used an L(T) function whose value at the absorption temperature is the same across all the wavelength ranges. Under our assumptions L(T) is given by:

$$L(T) = -\frac{1}{k_{eff}(T)} \ln\left(\frac{F_a}{F_b}\right)$$              Equation 3

where $k_{eff} = f(\text{HI}, T)k_{HI} + A_{He}[f(\text{HeI}, T)k_{HeI} + f(\text{HeII}, T)k_{HeII}]$     Equation 4

$k_{eff}$ is the effective absorption coefficient with unique temperature dependencies in the four wavelength ranges described earlier in this section. Respectively, $k_{HI}$, $k_{HeI}$, and $k_{HeII}$ are absorption cross-sections of $H^{1+}$, $He^{1+}$, and $He^{2+}$, which were calculated from the photo-ionization cross-sections from Verner et al. (1996), and the *f* functions are the temperature-dependent fractional abundances of each species which were taken from the CHIANTI database (Del Zanna et al. 2015; Dere et al. 1997) under the assumption of ionization equilibrium.

Using measurements of $\tau$ in the appropriate wavelength channels and the respective $k_{eff}$ functions, L(T) can be plotted versus temperature as shown in Figure 4. The intersection point of these curves gives values of the column density and the corresponding temperature of the absorbing plasma. Results from this diagnostic technique for the filament under consideration are described in Section 4.3.

While the helium abundance ($A_{He}$) is expected to vary within each filament, it was set as 5% for the purposes of this analysis. The diagnostic results using $A_{He}$ = 1% and 10%, compared with the results using $A_{He}$ = 5% show within an order of magnitude difference in column density and no change in temperature. This is expected because the temperature only depends on HeI and HeII, and changing $A_{He}$ only changes the point at which the L(T) crossing takes place vertically in Figure 4.

Since the Landi and Reale (2013) diagnostic technique used CHIANTI ion fractions, it implicitly assumes the plasma is in ionization equilibrium, which the CME plasma may not be in. A test and discussion of this assumption is presented in Section 8 of this paper.

### 4.2. Estimation of $F_b$:

In order to compute $F_b$ (Equation 2), reasonable approximations needed to be made based on the time-dependent geometry of the filament. One method to approximate the value of $F_b$ is to assign the measured intensity at a point close to the filament but outside of it (e.g. Gilbert et al. 2005). Since we considered 102 timestamps over the course of 50 minutes where the filament geometry changes considerably, we chose instead a relatively quiescent period before the flare and filament erupted when no transient structures were apparent. If the intensities at the locations of the plasma parcels (Figure 2) did not vary considerably, we could use them as a proxy for $F_b$.

Before the flare erupted at 03:45 UT, we monitored the intensities at each location of the plasma parcels from 03:00 UT to 03:44 UT using the 12 to 24 second cadence processed *SDO/AIA* images. The means, standard deviations of intensities at each location, and their intensity profiles over time in the 193 Å and 304 Å channels were analyzed. While there was some variability in the time-scales of the cadence of *AIA* observations, the intensities in the plasma parcels did not change significantly over timescales of a few minutes, and the trends in intensity remained intact over the 40 minutes analyzed. We concluded that the intensities at 03:40 UT were reasonable approximations for $F_b$ of the absorbing plasma along the three paths under scrutiny because (1) the mean intensity at this time was comparable to the mean values over the 40 minutes analyzed with reasonable standard deviations, and (2) the standard deviation of intensities from 03:00-03:40 UT for a majority of the plasma parcels analyzed was within 10% of the mean.

### 4.3. Results of Diagnostics:

In the diagnostics discussed here, the 335 Å channel was not considered due to the low signal-to-noise ratio in the data. The results of the diagnostic technique for one time-step along Path 1 are shown in Figure 4. Here [171 Å, 193 Å, 211 Å] and 304 Å belong to wavelength regions with different temperature functions of the effective absorption coefficient, enabling us to measure the column density and temperature of the 30x30 pixel plasma parcel at that time and location from the crossing of all the curves. The crossing point is well-defined and allowed for accurate diagnostics.

Diagnostic results such as those shown in Figure 4 were produced for 102 timestamps between 04:10 UT – 05:00 UT for all three plasma trajectories within the filament (Figure 2). Distinct intersections in the L(T) curves were obtained for approximately the first 30 minutes of analysis for all three plasma paths, allowing us to measure the column density and temperature at these times. These results are shown in Figure 5. The bottom x-axis for plots in Figure 5 show time, and the top x-axis shows the corresponding height of the filament (see Section 5). The left panels of Figure 5 show logarithm of the column density for paths 1, 2, and 3 respectively, and the right panels show logarithm of the temperature for paths 1, 2, and 3. Between ~4:40 UT and 5:00 UT the L(T) curves for passbands in the two wavelength domains either did not intersect or did so with high uncertainty. This can either be attributed to the temperature of the absorbing material falling below 15,000 K at which point the diagnostic technique does not work, or the absorption features were not strong enough consistently in all the wavelength channels.

## 5. Kinematics of the Filament Plasma in the Low Solar Corona:

In order to determine the height and speed evolutions of the plasma parcels as they travel along the three paths (Figure 2) as part of the erupting filament, *STEREO A EUVI, COR1*, and *COR2* observations were utilized. The *STEREO A* instrument suite was chosen over *STEREO B* since it gave us a marginally better view of the active region. The filament eruption is often identified as the brightest part of the CME in white-light images (Parenti et al. 2014) and since STEREO was in near quadrature with *SDO* at this time, the tip of the filament eruption seen in the time-series images was taken to roughly correspond to the height at which the plasma parcels within the filament were observed. All *SECCHI* images were first processed from level 0.5 to level 1.0 using the secchi_prep.pro[3] SolarSoft routine. The tip of the filament remained in the field of view of *EUVI* till ~4:10 UT after which we measured the height using *COR1*. Speed evolution approximations were determined from the height evolution profiles. Results of this analysis are shown in Figure 6.

Accurate estimates of the height of the plasma parcels within the filament were limited by the fact that this was a surge CME (Vourlidas et al. 2003, Vourlidas et al. 2017) with indistinguishable parts. This made it unfeasible to precisely locate the plasma observed in *AIA* to the corresponding *STEREO* plasma. The standard deviations in speed values presented in Figure 6 are hence considered in the computation of composition described in Section 7. All three filament plasma parcels analyzed were assumed to be at the same height and traveling at the same speed which may not be a realistic approximation, and thus is a limitation to be considered in interpreting the outcome of this analysis.

## 6. Neutral Hydrogen Density, Proton Density & Electron Density of the Filament Plasma in the Low Solar Corona:

The average lower limit to the neutral hydrogen density can be directly computed from the column density if the line of sight depth of the absorbing parcel is known, following the equation (Landi & Reale 2013):

$N_H = n_L / L$                               Equation 5

where $L$ is the depth of the absorbing plasma along the line of sight, $n_L$ is the column density, and $N_H$ is the neutral hydrogen density. Assuming homogenously absorbing plasma parcels, a unity filling factor is used. While previous studies (e.g. Landi & Reale 2013) have assumed arbitrary values of this depth, time series of the depth of the absorbing plasma are determined using *STEREO A/EUVI* and *COR 1* images that observe the filament eruption.

At each time stamp where *STEREO A/EUVI* and *COR1* images are available, the latitudes at which the plasma parcels are tracked along the three paths were recorded. The depth of the absorbing plasma at each of these latitudes was used as approximations of the depth along the line of sight at that instant. Profiles for line of sight depth of the plasma are shown in Figure 7. It is noted that there is little variation between the depth profiles for the three paths, and the values are large due to the large angular scale of the filament eruption. Appropriate fits to the data are chosen to match the column density measurements. This process is limited by the low cadence *EUVI* and *COR 1* observations, and the fact that at certain timestamps the filament line of sight was outside the field of views of *EUVI* and *COR 1* despite the overlap. Results for the neutral hydrogen density for paths 1, 2, and 3 are shown in Figure 8.

The proton density is computed using the read_ioneq.pro[4] CHIANTI code. Ratios of neutral hydrogen abundance to proton abundances are drawn under conditions of ionization equilibrium from temperature in the range of $10^4$ to $10^9$ Kelvins. Using these ratios and the neutral hydrogen density (Figure 8), proton density profiles are derived. As discussed in Section 4.1, an ionization equilibrium approximation is a critical caveat for these computations since this might not be a realistic approximation. This approximation is discussed in Section 8. Results for the proton number density, under these approximations, for Paths 1, 2, and 3 are shown in Figure 9.

The electron number density was then computed from the proton number density using proton_dens.pro[5] SolarSoft code. Temperatures measured and reported in Section 4.3 were used as inputs to the code to compute the ratio of proton density to electron density using abundances and ion balance files.

# 7. Ionization History of the Filament Plasma in the Low Solar Corona:

In order to determine the ionization history of the plasma parcels we used the Michigan Ionization Code (MIC), an ion composition model that predicts the evolution of the ion abundances of wind plasma leaving the Sun from the source region out through the freeze-in point and beyond to Earth. MIC was developed and used in Landi et al. (2012a). It solves for ionization and recombination for a chosen element as it travels outward from the Sun using the following set of equations:

$$\frac{\partial y_m}{\partial t} = n_e[y_{m-1}C_{m-1}(T_e) + y_{m+1}R_{m+1}(T_e)] + y_{m-1}P_{m-1} - y_m[n_e(C_m(T_e) + R_m(T_e)) + P_m]$$

Equation 6

$$\Sigma_m y_m = 1,$$ 

Equation 7

where the photoionization term, $P_m$, is given by

$$P_m = \int_{v_m}^{\infty} \frac{4\pi J(v)\sigma_m(v)}{hv} dv$$ 

Equation 8

where $T_e$ is the electron temperature, $n_e$ is the electron density, $R_i$ and $C_i$ are the total recombination and ionization rate coefficients, $y_m$ is the fraction of the element X in charge state $m$, $h$ is the Planck constant, $c$ is the speed of light, $\sigma_m$ is the photoionization cross section for the ion $m$, $v_m$ is the frequency corresponding the ion's ionization energy, and $J(v)$ is the average spectral radiance of the Sun at frequency $v$. Equations 6 and 7 for each element are solved numerically as a set of stiff ordinary differential equations using a fourth-order Runge-Kutta method (Press et al. 2002). To ensure computational efficiency, the step size is set adaptively and the accuracy of the integrator is tested to ensure high accuracy.

MIC was run for the plasma parcels within the filament for Paths 1, 2, and 3. The effects of photoionization on the charge state composition of the filament plasma was taken into account following the discussion in Landi & Lepri (2015), where they found that photoionization had significant effects on the charge state distribution of C, N, and O. They also documented significant effects on Fe charge states in CMEs. MIC also allowed us to account for the impact of the energetic M-class flare that erupted prior to the filament liftoff on the ionization states. The flare spectrum for an X-class flare is used in the computation presented here. Inputs to the MIC were as follows:

1. *Electron temperature* - values described in Section 4.3 and shown in Figure 5.

2. *Bulk speed of plasma* - An ionization profile was produced using the polynomial fits to the (mean values + standard deviations) of the speeds, and another was produced using the polynomial fits to the (mean values – standard deviations) of the speeds (Figure 6, right). This was done to account for the uncertainty in the speed determination. However, results from both sets of speeds did not show significant variation hence the results using the mean value of speeds are presented in this report.

3. *Electron density* - values described in Section 5.

Results of the MIC are shown in three sets of plots in Figure 10. The top set of plots is for Path 1, the middle set is for Path 2, and the bottom set is for Path 3. The bottom panel in each set shows the relative abundance of the charge states of C, N, O, and Fe to the total abundance of their respective elements, represented with color bars, as the plasma parcels within the filament travel from 1.6 – 2.4 solar radii. The top panel in each set shows the relative abundance of the charge states of C, N, O, and Fe to the total abundance of their respective elements at the maximum height of the plasma parcels along the measured trajectories. These plots were produced to mimic the analysis in Gruesbeck et al. (2011) to compare with in-situ values to test the freeze-in condition.

# 8. Discussion:

## 8.1. Ionization Equilibrium:

The absorption diagnostic technique (Section 4.1) was implemented under the assumption of ionization equilibrium since the ion fractions used from CHIANTI are computed under this assumption. Additionally, in Section 6 we described the proton density calculations from the neutral hydrogen density using read_ioneq.pro, which also assumes ionization equilibrium. In order to test the validity of this approximation, we compared ionization profiles of $H^{1+}$, $He^{1+}$, and $He^{2+}$ described in Section 7 with the equilibrium profiles of the same ions along the three plasma parcel trajectories. We found the departure from equilibrium was within 10% for the three ions, justifying the approximation under which our calculations were performed.

## 8.2. Emission in 304 Å:

Previous studies (e.g. William et al. 2013, Lee et al. 2017) discounted the 304 Å passband for absorption diagnostics due to significant emission contributions from the erupting filament itself. Since the Landi and Reale (2013) absorption diagnostic technique is based on the assumption that the contribution from emission in all channels are negligible, it is important to quantify the emission contribution in 304 Å for the filament under consideration. This is done in two steps: (1) computing the filling factor of the filament plasma using the EUVI 304 Å channel data where the filament is observed entirely in emission, and (2) comparing the *AIA* count rates predicted using this filling factor with the measured count rates for a plasma parcel within the filament. The plasma parcel tracked along Path 1 in *AIA* at 04:16 UT was used to make these estimates, since *EUVI* observations coincident with *AIA* observations were available at this time with strong counts. The flux of photons expected to be incident at the *EUVI* detector on *STEREO B* is given by:

$$F_{STEREO-B/EUVI} = \frac{1}{4\pi d^2} N_e^2 G(T) A L = 1.57 * 10^5 \; photons.cm^{-2}s^{-1} \quad \quad \text{Equation 9}$$

Here, $d$ is the distance between the Sun and *STEREO B* = 1.04 AU, $N_e$ is the electron density = $1.79*10^{11}$ cm$^{-3}$, $G(T)$ is the contribution function from CHIANTI = $1.15*10^{-16}$ photons/cm$^{-3}$s$^{1}$, $A$ is the area of an EUVI pixel = $7.5*10^7$ cm, and $L$ is the line of sight depth of the filament plasma in EUVI. $L$ is approximated as the width of the absorbing plasma in *AIA* at that time step = $8.8*10^9$ cm. A calibration factor from the get_calfac.pro[6] SolarSoft routine is used. Using an effective area of 0.0505 cm$^2$ for the *EUVI* detector (values distributed by the *SECCHI* team in SolarSoft), we estimated an *EUVI* flux of 8080 photons/second. The measured *EUVI* fluxes at locations at the latitude of the *AIA* plasma parcel were ~ 0.8 – 1.75 photons/second, giving a filling factor of the filament plasma ~ (1.05 – 1.73) * 10$^{-4}$.

Equation 9 was similarly used to compute the predicted flux incident on the *AIA* detector, using $A = 1.96*10^{15}$ cm$^2$, $L = 0.321$ solar radii (at the altitude the plasma parcel was observed in *AIA*), $N_e = 1.79*10^{11}$ cm$^{-3}$, and $G(T) = 1.15*10^{-16}$ photons/cm$^{-3}$s$^{1}$. This gave a value of $F_{SDO/AIA} = 5.78*10^4 \; photons.cm^{-2}s^{-1}$. Using a calibration factor from get_calfac.pro, the predicted flux at *AIA* in 304 Å ~ 2196.4 DN/s. Using the filling factor from the *EUVI* 304 Å passband, this gave us *AIA* 304 Å count rates for the filament plasma ~ 0.3 DN/s, which is the expected contribution from emission. This is 10% of the flux observed by *AIA* in the 30x30 pixel path 1 plasma parcel tracked at 04:16 UT. In the current investigation, we present the emission contribution in the 304 Å channel as a critical caveat to our analysis. Its contribution is fairly small but could still alter the diagnostics of temperature. The presence of this emission slightly increases the temperature of the crossing point in plots such as those in Figure 4 but does not change the column density significantly since the L(T) values for the other channels are almost constant for large temperature changes. The Landi and Reale (2013) diagnostic technique does not take into account the presence of emission contribution, but we are extending this technique to account for it and will report our findings in subsequent work (Kocher et al. 2018, in preparation). For the purposes of this particular ICME, the small effect that emission has on the temperature measurements does not invalidate our results.

### *8.3. Freeze-in Condition in Filament Plasma:*

The disturbance from the ICME associated with the filament eruption under scrutiny was registered by the RC list at 17:51 UT on August $5^{th}$, 2011, with the ICME duration from 22:00 UT, August $6^{th}$, 2011 – 22:00 UT, August $7^{th}$, 2011. The heavy ion charge states during the ICME period and in the 12 hours before and after were analyzed. No signatures of cold plasma in the form of low charge states of carbon, nitrogen, oxygen, and iron were recorded in the in-situ observations at 1 AU by *ACE*, following the criteria outlined in Lepri and Zurbuchen (2010). Since filamentary signatures were not detected, we were unable to validate the 5% He/H abundance value assumed to calculate the effective absorption coefficient. This lack of low-ionization plasma in the measurements could be due to two causes. First, the filament plasma was not frozen in at the maximum height (~2.4 solar radii) that we've been able to conduct our analysis. The plasma may have continued to ionize due to additional heating at larger heights. Conversely, if the plasma continued to cool, we would expect most of the material to end up in neutral state, in which case *ACE* would not be able to detect the filament composition due to lack of neutral particle sensors onboard, despite incidence. The measured high-density of the filament plasma also supports the fact that the plasma is not frozen in at this altitude. Second, the lack of low ionization states in in-situ measurements of the CME plasma could also be a direct measure of the low filling factor of the filamentary material within the ICME. In fact, even though ~70% of all CMEs (e.g. Gopalswamy et al. 2003) are associated with filament eruptions, only ~4% of all CMEs incident at L1 have signatures of cold filament material (Lepri and Zurbuchen 2010). Our estimate of a filling factor of ~$10^{-4}$ for the cold absorbing filament plasma corroborates this scenario. To connect the coronal charge states of the core plasma with in-situ values, the filament eruption needs to be analyzed in observations at larger heights, or a different CME with in-situ filament signatures needs to be studied. This will be addressed in a companion study utilizing observations of the CME further out in the solar corona.

### *8.4. Heterogeneous & Heated Plasma:*

The temperature and density measurements (Figures 5, 8, 9) indicate that the plasma parcels within the filament eruption traveling along the three paths (Figure 2) behaved differently; thus, the filament plasma was heterogeneous. Likewise, in their analysis of the *Hinode/EIS* spectra of a CME core, Landi et al. (2010) concluded the core was made of two different plasma components moving coherently but with different temperature, filling factors and densities.

Our density and temperature measurements were compared with density and temperature profiles for an adiabatically expanding plasma parcel using the initial temperature and density we measured for each parcel trajectory as boundary conditions. This comparison provided strong evidence for the presence of additional sources of heating throughout the observed filament trajectory because our measured density and temperature profiles did not decrease as rapidly or in a manner similar to an adiabatically expanding plasma. Our measured density profiles stayed constant or decreased at a gradual rate for a majority of the filament plasma trajectory and ended at values similar to those expected for an adiabatically expanding plasma. However, the electron temperature profiles did not fall to values expected for an adiabatically expanding plasma and either remained mostly constant or gradually decreased for the three trajectories. The filament plasma had to be continuously heated throughout its trajectory, also leaving the question open as to whether the erupting filament is still subject to additional heating beyond the heights observed by *AIA* critical to our understanding of CME dynamics. This question will be addressed in a companion study.

### *8.5. Energetics of Filament Eruption:*

Initial estimates of the kinetic energy, gravitational potential energy, and thermal energy of the plasma parcels along the three trajectories were made per unit volume. The kinetic energy, gravitational potential energy, and thermal energy of the plasma parcels along Paths 1 and 2 within the filament displayed similar trends with localized fluctuations for the first half of the evolution, followed by sharp drops. The kinetic energy, gravitational potential energy, and thermal energy for Path 3 on the other hand displayed an initial increase followed by a sharp drop. The ratio of kinetic energy per unit volume to thermal energy per unit volume vary between 28 to 200, and the ratio of thermal energy per unit volume to gravitational potential energy per unit volume vary between 0.002 to 0.02. These trends indicated that the gravitational potential energy was greater than the kinetic energy, and the thermal energy was the less than both the gravitational potential energy and kinetic energy, for the entire evolution of the plasma parcels along the three trajectories. The lower limit of our kinetic energy/thermal energy estimates, and the upper limit of our thermal energy/gravitational potential energy estimates were in good agreement with those reported by Landi et al. (2010) for their CME core at 1.9 solar radii. A more detailed analysis of the energy budget of the CME core in the low solar corona will follow in the companion paper, after detailed computations of the time evolution of the geometry of the erupting filament.

## 9. Conclusion:

In order to constrain the physics of a filament eruption low in the solar corona and the partition of thermal and kinetic energy of their plasmas, we determined the evolution of the thermal and kinetic properties of CME cores using a combination of remote sensing and in-situ observations. Three 30x30 pixel square plasma parcels within an absorbing filament eruption were tracked along three possible paths from 04:10 UT – 05:00 UT in multi-wavelength *SDO/AIA* images on August $4^{th}$, 2011. The column density and electron temperature of the plasma were measured with high accuracy using the absorption diagnostic technique outlined in Landi & Reale (2013). The kinematics of the filament eruption were measured using *STEREO* multi-wavelength images and coronagraphs, and estimates of the line of sight length of the plasma from these images allowed for the calculation of the neutral hydrogen density. The proton and electron densities calculated from the neutral hydrogen density, along with the temperature and speed estimates, served as inputs to the Michigan Ionization Code to study the time and spatial evolution of the ionization status of the CME core plasma. The charge states within the heterogeneous core plasma were not frozen-in yet, prompting us to continue this investigation using instrument observing the filament eruption further out in the solar corona. The thermodynamic, kinematic, composition, and energy profiles reported in this paper represent high-cadence measurements of the evolution of the expanding filamentary plasma that was closely tracked in the low solar corona. These measurements can be powerful tools to constrain existing models of CME eruptions, and determine the physical processes of heating and acceleration that shape the eruptions at their source and throughout the low solar corona. We found the filament plasma to be nonhomogeneous, continuously heated throughout its trajectory, and its energetics favored gravitational potential energy over kinetic or thermal energy. Future studies will address the question of heating of the filament plasma in the low solar corona with a complete discussion of the evolution of the filament geometry, heating and dynamics of the filament plasma at heights larger than those observed in *AIA*, and compare our measurements to models of CME formation and evolution. Additionally, the Landi & Reale (2013) absorption diagnostic technique will be adapted to allow use of channels that have significant emission contributions, allowing for wider applications of the technique.

## 10. Acknowledgements:

The work of M.K. is supported by NESSF grant NNX16AP03H, the work of S.T.L. is supported by NASA grant N00173141G904, NSF grant 1460170. The work of E.L. is supported by NNX15AB73G, AGS-1358268. We offer our deepest appreciation to Dr. Angelos Vourlidas and Dr. Benjamin Lynch, who offered valuable insight during various stages of this research.

# 11. References:

Gruesbeck, J. R., Lepri, S. T., Zurbuchen, T. H., & Antiochos, S. K. 2011, ApJ, 730, 103

Landi, E., Gruesbeck, J. R., Lepri, S. T., Zurbuchen, T. H., & Fisk, L. A. 2012, ApJ, 761, 48

Rodkin, D., Goryaev, F., Pagano, P., et al. 2017, Sol. Phys., 292, 90

Geiss, J., Gloeckler, G., von Steiger, R., et al. 1995, Science, 268, 1033

Howard, R. A., Sheeley, N. R., Jr., Michels, D. J., & Koomen, M. J. 1985, JGR, 90, 8173

Illing, R. M. E., Hundhausen, A. J. 1985, JGR, 90, 275

Hundhausen, A. J. 1987, Sixth International Solar Wind Conference, 181

Hundhausen, A. 1999, The many faces of the sun: a summary of the results from NASA's Solar Maximum Mission., 148

Howard, R. A., Moses, J. D., Vourlidas, A., et al. 2008, SSR, 136, 67

Kaiser, M. L., Kucera, T. A., Davila, J. M., et al. 2008, SSR, 136, 5

Lemen, J. R., Title, A. M., Akin, D. J., et al. 2012, Sol. Phys., 275, 17

Pesnell, W. D., Thompson, B. J., & Chamberlin, P. C. 2012, Sol. Phys., 275, 3

House, L. L., Illing, R. M. E., Sawyer, C., & Wagner, W. J. 1981, BAAS, 13, 862

Webb, D. F., & Hundhausen, A. J., 1987, Sol. Phys., 108, 383

Gilbert, H. R., Holzer, T. E., Burkepile, J. T., & Hundhausen, A. J. 2000, ApJ, 537, 503

Zhang, J., Dere, K. P., Howard, R. A., Kundu, M. R., & White, S. M. 2001, ApJ, 559, 452

Gopalswamy, N., Lara, A., Yashiro, S., Nunes, S., & Howard, R. A. 2003, Solar Variability as an Input to the Earth's Environment, 535, 403

Landi, E., Raymond, J. C., Miralles, M. P., & Hara, H. 2010, SOHO-23: Understanding a Peculiar Solar Minimum, 428, 201

Chen, P.F. 2011, Living Rev. Sol. Phys., 8:1

Gilbert, H. R., Holzer, T. E., MacQueen, R. M. 2005, ApJ, 618, 524

Gilbert, H., Kilper, G., Alexander, D., & Kucera, T. 2011, ApJ, 727, 25

Landi, E., Reale, F. 2013, ApJ, 772, 71

Landi, E., Miralles, M. P. 2014, ApJL, 780, L7

Hannah, I. G., & Kontar, E. P. 2012, AAP, 539, A146

Hannah, I. G., & Kontar, E. P. 2013, AAP, 553, A10

Lee, J. -Y., Raymond, J. C., Reeves, K. K., Moon, Y. J., & Kim, K. -S. 2017, ApJ, 844, 3

Domingo, V., Fleck, B., & Poland, A. I. 1995, Sol. Phys., 162, 1

Hundhausen, A. J., Gilbert, H. E., & Bame, S. J. 1968, JGR, 73, 5485

Hundhausen, A. J. 1972, Physics and Chemistry in Space, 5,

Bame, S. J., Asbridge, J. R., Feldman, W. C., & Kearney, P. D. 1974, Sol. Phys., 35, 137

Buergi, A., & Geiss, J. 1986, Sol. Phys., 103, 347

Gloeckler, G., Cain, J., Ipavich, F. M., et al. 1998, SSR, 86, 497

Lepri, S. T., Zurbuchen, T. H., Fisk, L. A., et al. 2001, JGR, 106, 29231

Zurbuchen, T. H., & Richardson, I. G. 2006, SSR, 123, 31

Richardson, I. G., & Cane, H. V. 2004, Journal of Geophysical Research (Space Physics), 109, A09104

Richardson, I. G., & Cane, H. V. 2010, Sol. Phys., 264, 189

Kocher, M., Lepri, S.T., Landi, E., Zhao, L., & Manchester, W. B., IV 2017, ApJ, 834, 147

Lynch, B. J., Antiochos, S. K., DeVore, C. R., Luhmann, J. G., & Zurbuchen, T. H. 2008, ApJ, 683, 1192

Reeves, K. K., Linker, J. A., Mikic, Z., & Forbes, T. G. 2010, ApJ, 721, 1547

Brueckner, G. E., Howard, R. A., Koomen, M. J., et al. 1995, Sol. Phys., 162, 357

Del Zanna, G., Dere, K. P., Young, P. R., Landi, E., & Mason, H. E. 2015, AAP, 582, A56

Dere, K. P., Landi, E., Mason, H. E., Monsignori Fossi, B. C., & Young, P. R. 1997, VizieR Online Data Catalog, 412

Lepri, S. T., & Zurbuchen, T. H. 2010, ApJL, 723, L22

Parenti, S., & Vial, J.-C. 2014, Nature of Prominences and their Role in Space Weather, 300, 69

Landi, E., Gruesbeck, J. R., Lepri, S. T., Zurbuchen, T. H., & Fisk, L. A. 2012, ApJl, 758, L21

Press, W.~H., Teukolsky, S.~A., Vetterling, W.~T., \& Flannery, B.~P.\ 2002, Numerical recipes in C++ : the art of scientific computing by William H.~Press.~xxviii, 1,002 p.~: ill.~; 26 cm.~ Includes bibliographical references and index.~ISBN : 0521750334,

Landi, E., & Lepri, S. T. 2015, ApJL, 812, L28

Vourlidas, A., Wu, S. T., Wang, A. H., Subramanian, P. & Howard, R. A. 2003, ApJ, 598, 1392

Vourlidas, A., Balmaceda, L. A., Stenborg, G. & Lago, A. D. 2017, ApJ, 838, 141

Jian, L. K., MacNeice, P. J., Taktakishvili, A., et al. 2015, Space Weather, 13, 316

Verner, D. A., Ferland, G. J., Korista, K. T., & Yakovlev, D. G. 1996, ApJ, 465, 487

Lee, J. Y., Raymond, J. C., Reeves, K. K., Moon, Y. J., & Kim, K. S., ApJ, 798, 106

Williams, D. R., Baker, D., & van Driel-Gesztelyi, L. 2013, ApJ, 764, 165

**Links to superscripts in manuscript:**

[1] http://www.lmsal.com/get_aia_data/

[2] http://www.heliodocs.com/php/xdoc_print.php?file=$SSW/sdo/aia/idl/calibration/aia_prep.pro

[3] http://www.heliodocs.com/php/xdoc_print.php?file=$SSW/stereo/secchi/idl/prep/secchi_prep.pro

[4] http://sao-ftp.harvard.edu/PINTofALE/pro/external/read_ioneq.pro

[5] https://hesperia.gsfc.nasa.gov/ssw/packages/chianti/idl/utilities/proton_dens.pro

[6] http://www.heliodocs.com/php/xdoc_print.php?file=$SSW/stereo/secchi/idl/prep/get_calfac.pro

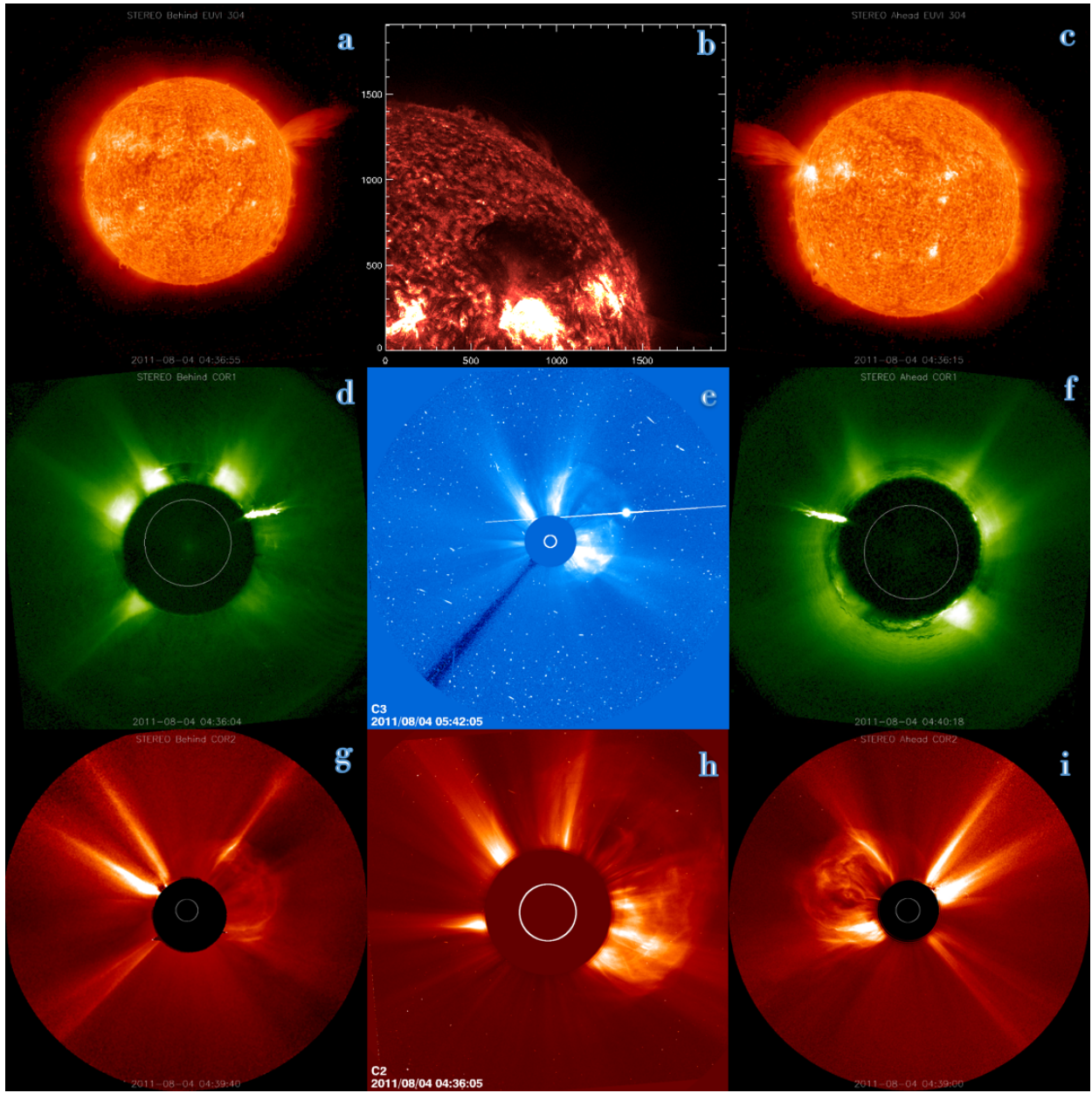

**Figure 1:** Multi-instrument observations of a ICME disturbance detected in-situ by *ACE/SWICS* at 17:51 UT, August 5[th], 2011 and reported by the RC list. a: *STEREO B/EUVI* 304 Å, b: *SDO/AIA* 304 Å, c: *STEREO A/EUVI* 304 Å, d: *STEREO B/COR 1*, e: *LASCO/C3*, f: *STEREO A/COR 1*, g: *STEREO B/COR 2*, h: *LASCO/C2*, i: *STEREO A/COR 2*

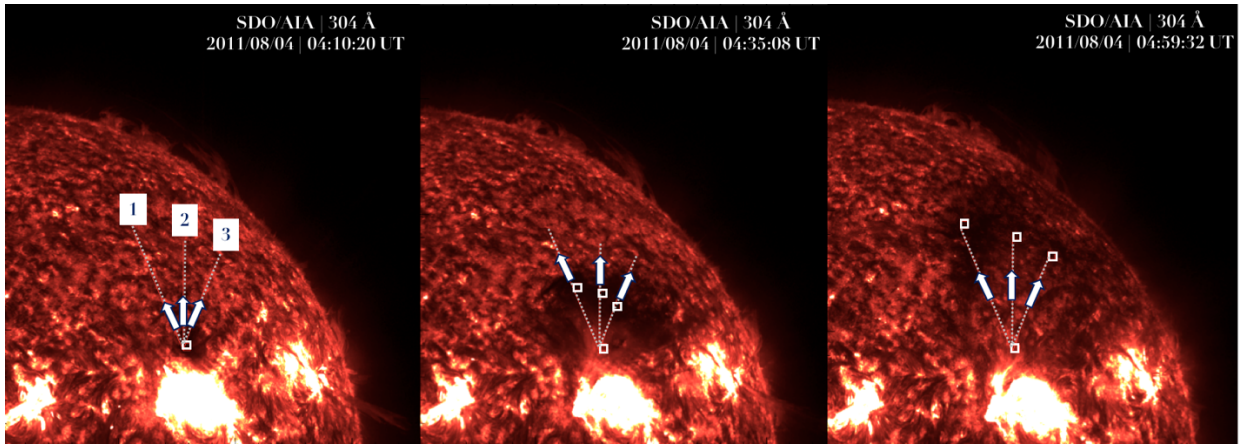

**Figure 2:** 304 Å *SDO/AIA* images of the August 4th, 2011 filament eruption at 04:10:20 UT (left), 04:35:08 UT (middle), and 04:59:32 UT (right). The dotted white lines 1, 2, and 3 represent the paths chosen to describe possible trajectories of the plasma within the white box (30x30 pixel) as the filament travelled through the low solar corona.

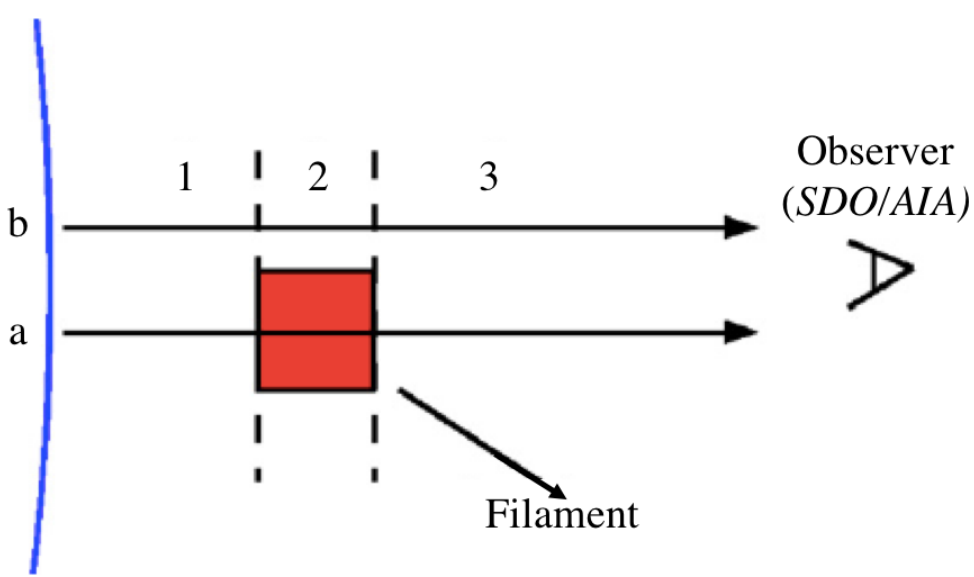

**Figure 3:** Schematic showing a simplified version of the geometry described in Section 4.1, when the absorbing plasma is against the solar disk. Regions 1, 2, and 3 are the 'background', 'filament' (absorbing plasma), and 'foreground' regions respectively for lines of sight *a* and *b*. Adapted from Landi & Reale (2013).

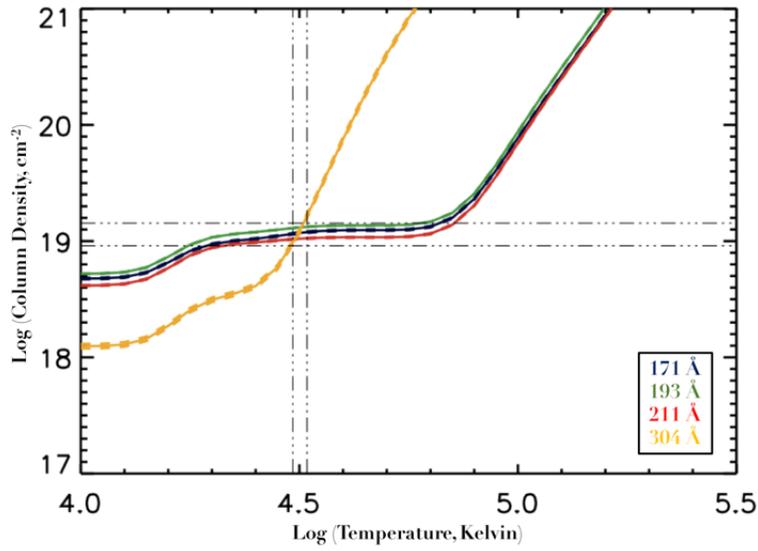

**Figure 4:** Diagnostic results for the absorbing plasma parcel location at 04:23 UT along Path 1. The y-axis is the L(T) function or the column density (cm$^{-2}$) and the x-axis is the temperature (K). The point where the L(T) curves for 171 Å, 193 Å, 211 Å, 304 Å intersect gives the measure of the column density and temperature of the absorbing plasma at that time and location.

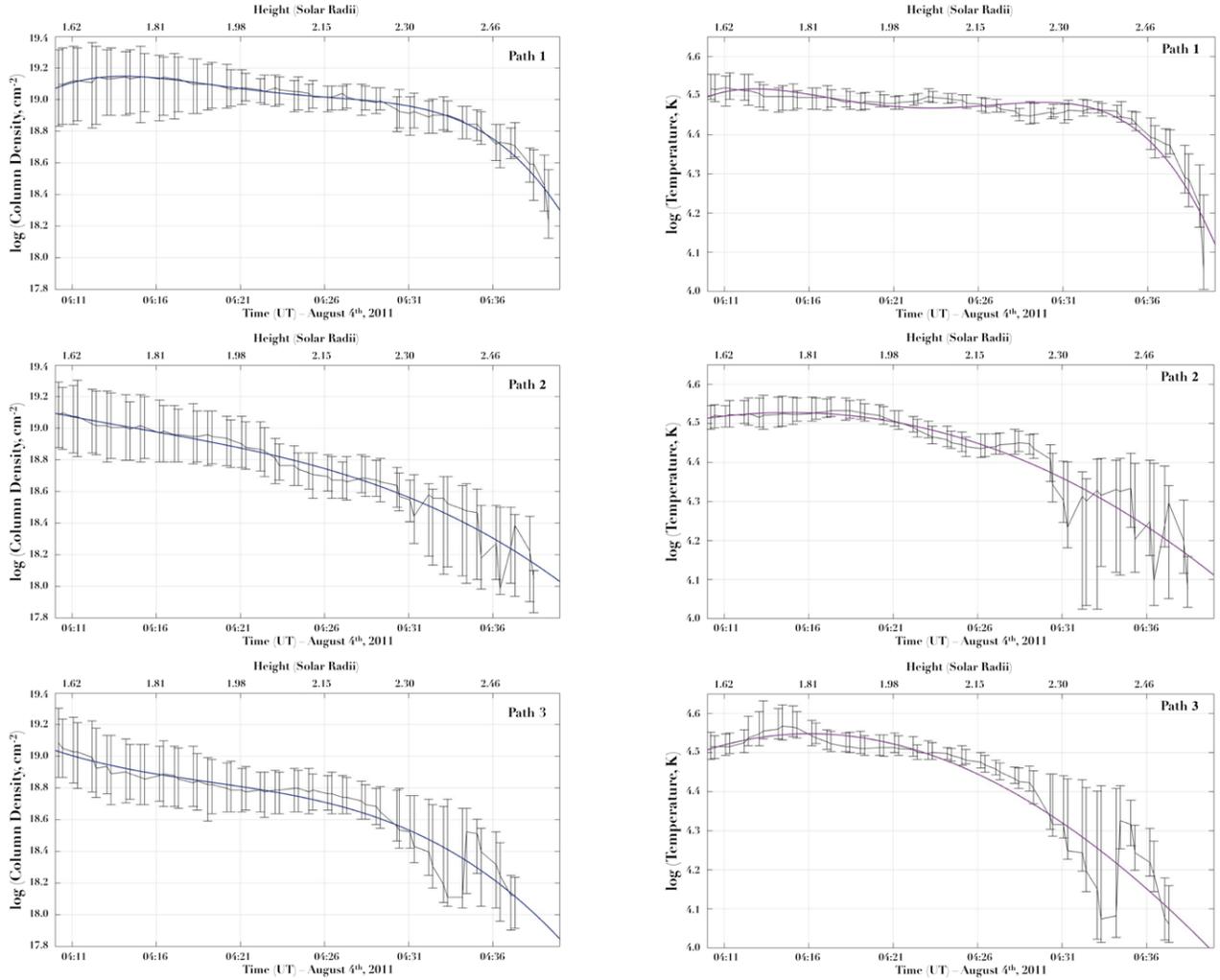

**Figure 5:** (left column) y-axis shows logarithm of column density (cm$^{-2}$) versus time (UT) on the bottom x-axis and the corresponding approximate height (solar radii) of the filament plasma parcels on the top x-axis, for paths 1, 2, and 3. (right column) y-axis shows logarithm of temperature (Kelvin) versus time (UT) on the bottom x-axis and the corresponding approximate height (solar radii) of the filament plasma parcels on the top x-axis, for paths 1, 2, and 3. The colored lines represent fits to the data.

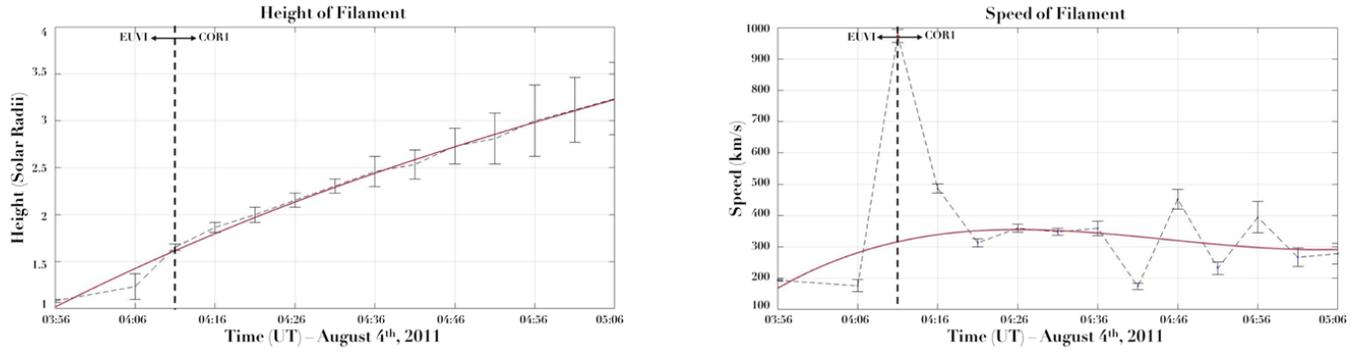

**Figure 6:** (left) plot of height of the filament eruption as seen by *STEREO A EUVI* and *COR1* versus time. (right) plot of speed of the filament eruption derived from the height vs time measurements. In the speed/time plot, the measurement at 04:11 UT is treated as an outlier for fitting purposes.

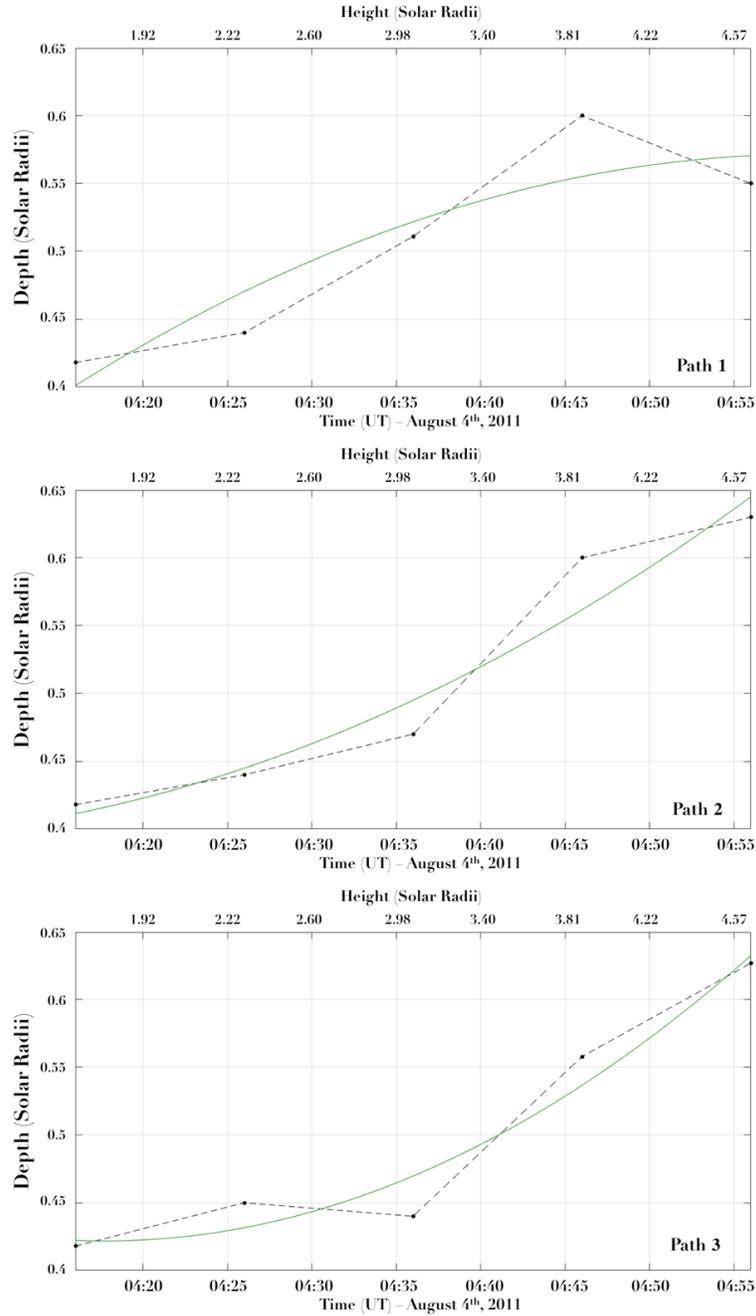

**Figure 7:** Depth of the absorbing plasma (solar radii) versus time (UT) on the bottom x-axis and the corresponding approximate height (solar radii) of the filament plasma parcels on the top x-axis, for Path's 1 (top), 2 (middle), and 3 (bottom) computed using *SDO/AIA*, *STEREO A/EUVI and COR1* images. The green curves are fits to the data.

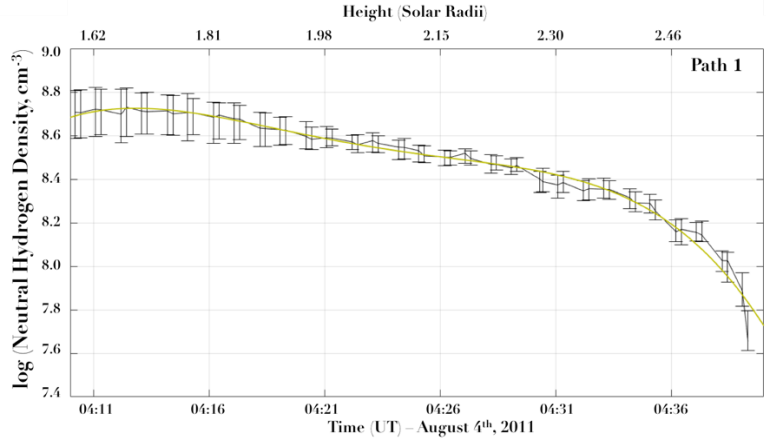

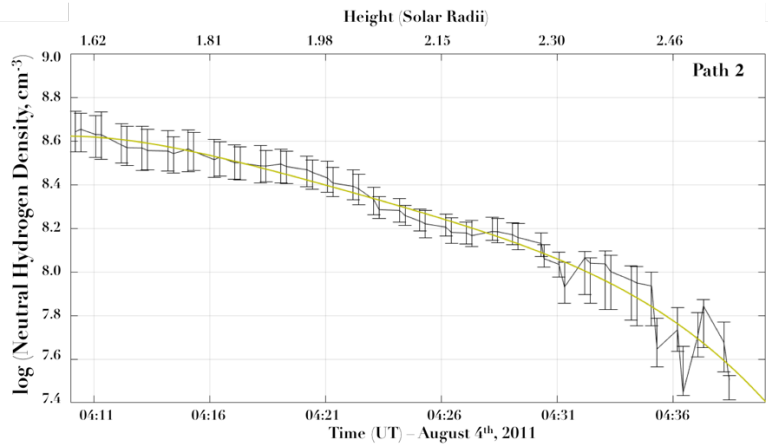

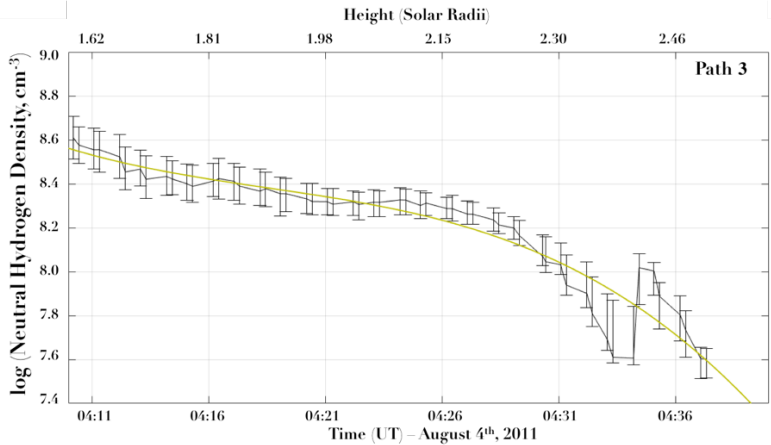

**Figure 8:** Number density of neutral hydrogen (cm$^{-3}$) versus time (UT) on the bottom x-axis and the corresponding approximate height (solar radii) of the filament plasma parcels on the top x-axis, for paths 1 (top), 2 (middle), and 3 (bottom), computed using Equation 4 and line of sight depth profiles shown in Figure 7. The orange curves are fits to the data.

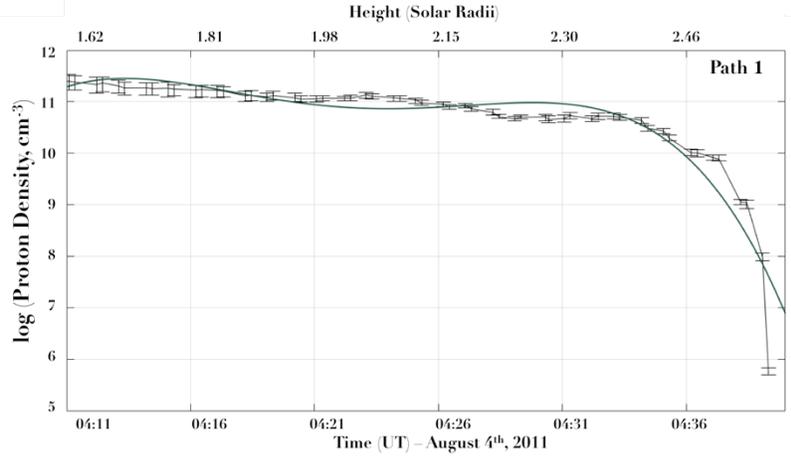
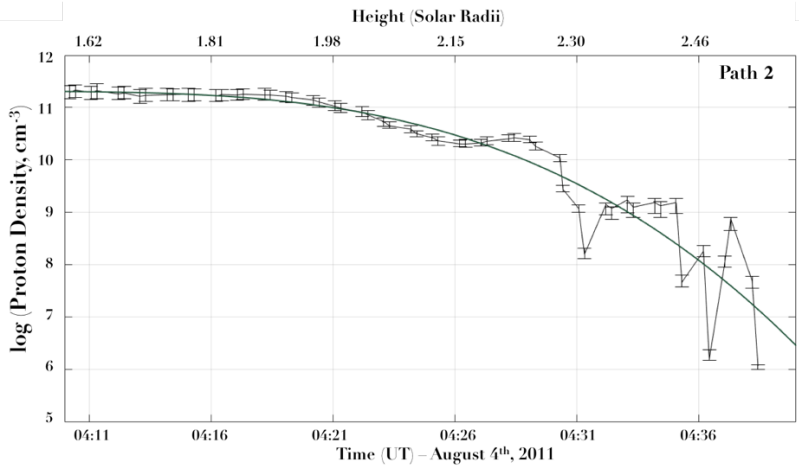
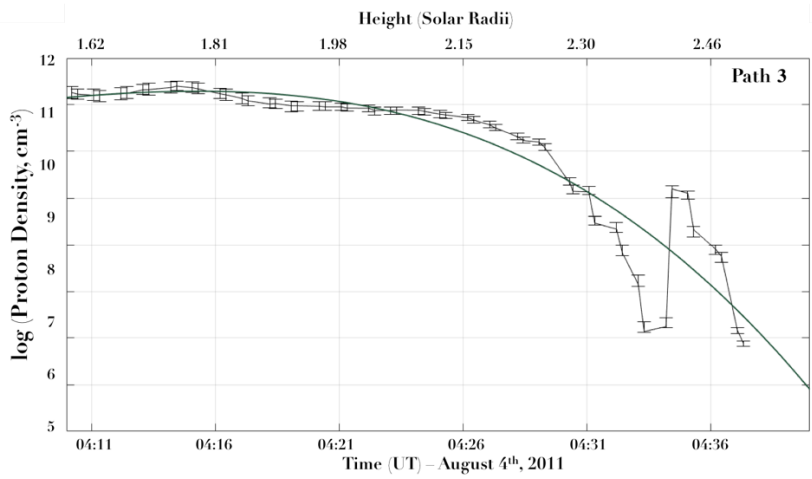

**Figure 9:** Proton number density (cm$^{-3}$) versus time (UT) on the bottom x-axis and the corresponding approximate height (solar radii) of the filament plasma parcels on the top x-axis, for paths 1 (top), 2 (middle), and 3 (bottom). The green curves are fits to the data.

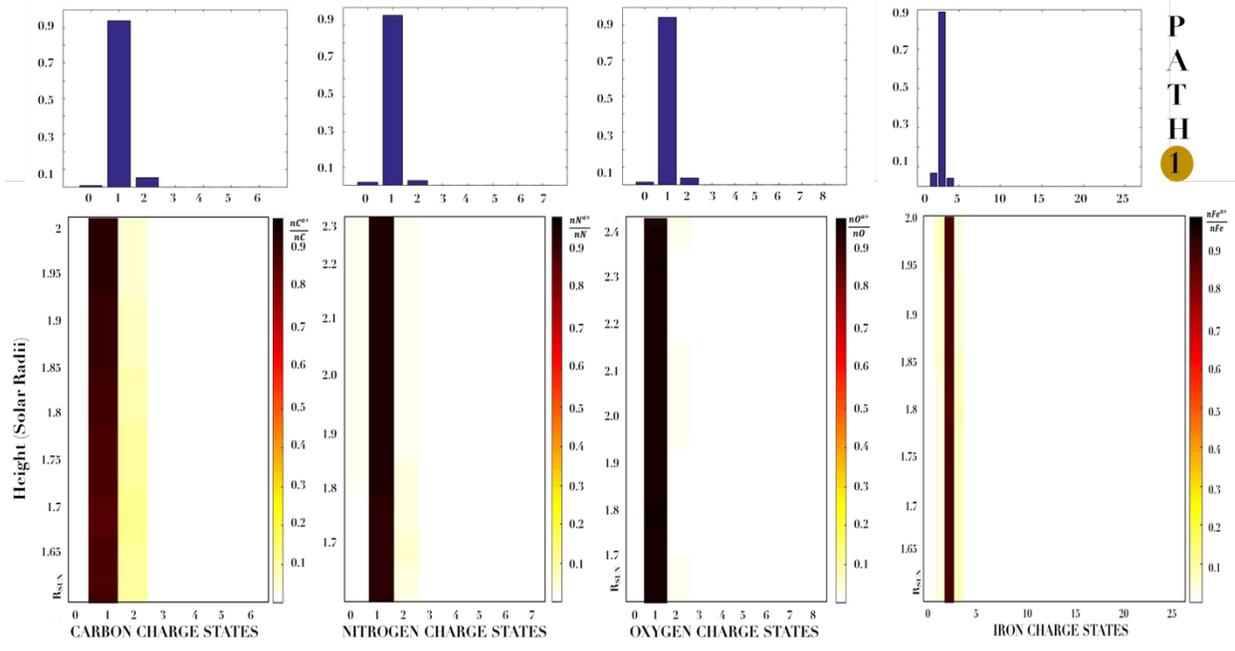

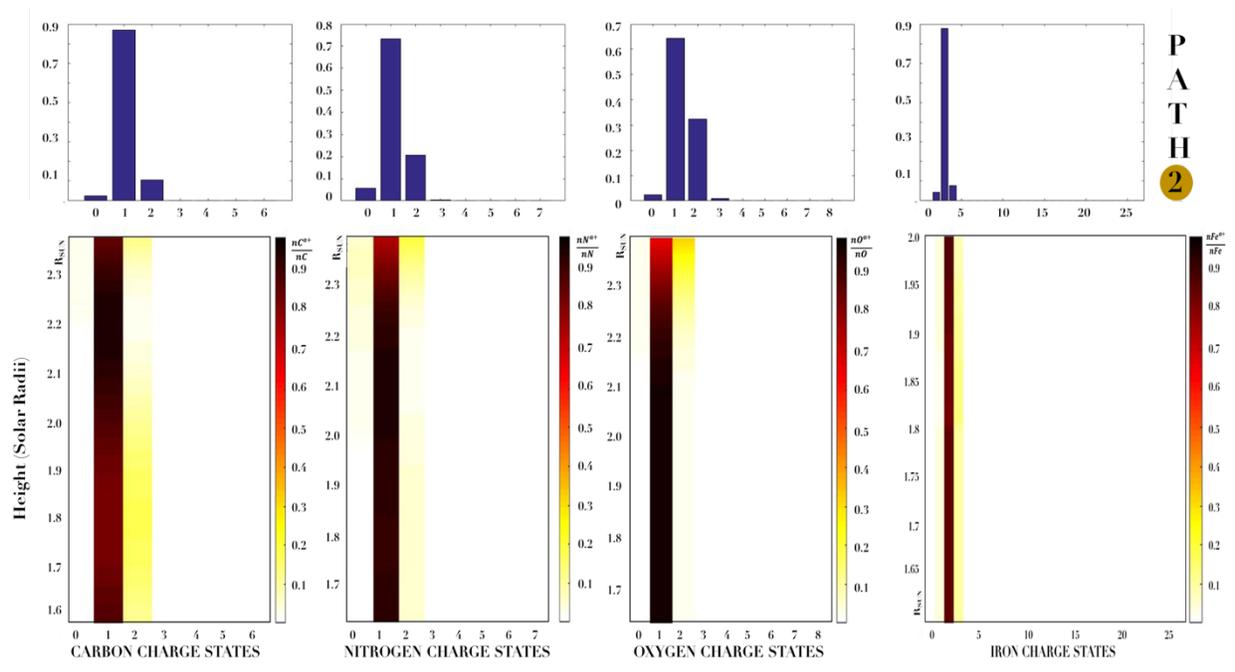

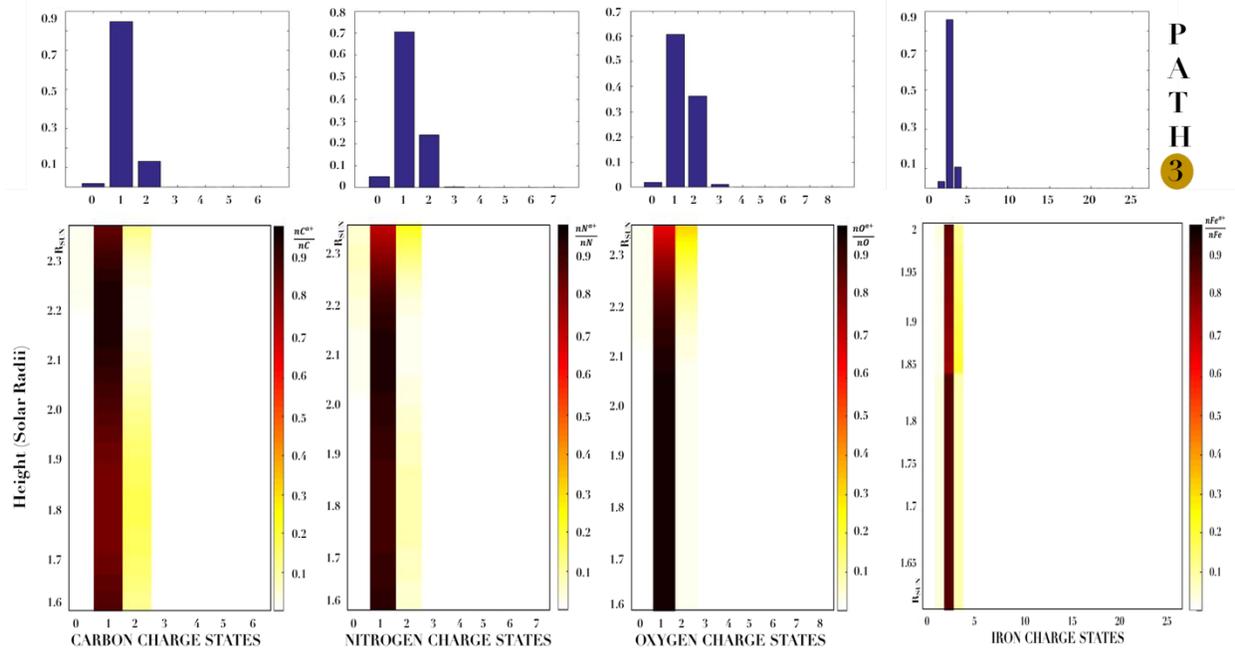

**Figure 10:** Results from Michigan Ionization Code using the measured values of density, temperature & speed for Path's 1(top), 2(middle), and 3(bottom). For each path: the bottom set of panels show the evolution of the charge states of Carbon, Nitrogen, Oxygen, and Iron with height, and the top set of panels show the final charge states observed at the maximum height. maximum height.